\documentclass[12pt,twoside]{article}
\usepackage{fleqn,espcrc1}
\usepackage{graphicx}
\usepackage[figuresright]{rotating}

\newcommand{\AmS}{{\protect\the\textfont2
  A\kern-.1667em\lower.5ex\hbox{M}\kern-.125emS}}

\def\dm2{\Delta m^2}
\def\sq2{sin^2(2\Theta)}
\def\nubar{\overline {\nu} }

\title{Progresses in the validation of the FLUKA atmospheric $\nu$ flux
calculation} 

\author{G. Battistoni\address[INFNMi]{INFN and Universit\`a di
Milano,  Dipartimento di Fisica, Milano, 20133, Italy}%
, A. Ferrari\address[CERN]{CERN, Geneva 23, Switzerland},
T. Montaruli\address[INFNBa]{INFN and Universit\`a di
Bari,  Dipartimento di Fisica, Bari, 70126, Italy}
and P.R. Sala\addressmark[CERN]}
       
\begin{document}

\maketitle

\begin{abstract}
The
FLUKA calculation of the 
atmospheric neutrino fluxes have been cross--checked
by comparing predictions on lepton fluxes
in atmosphere to experimental data. 
The dependence of predicted 
neutrino fluxes on the 
shape and normalization of primary spectrum is also investigated.
\end{abstract}

\section{Introduction}
The two essential features of the atmospheric neutrino
fluxes, namely the up-down symmetry and the $\mu/e$ ratio, are
now considered to be largely independent from the calculation
scheme\cite{now2000}. 
Theoretical uncertainties should still have an impact, together
with experimental systematics, in the 
determination of the relevant physical parameters of neutrino
oscillations\cite{lip2001}. However, the Super-Kamiokande allowed
region in the $\dm2-\sq2$ space remains more or less invariant\cite{Totsuka}
when the Monte Carlo predictions change from the original HKKM
results\cite{hkkm} to the FLUKA ones\cite{fluka}, despite the fact that the
two set of predictions have important differences. The most significant
one is the normalization of the flux, which, for
the same neutrino cross--sections, gives a 15$\div$20\% lower event
rate in the case of FLUKA. The same ratio approximately exists
between FLUKA and the original Bartol flux\cite{bartol} for the same
primary flux. 
This difference is due to  the features of the hadronic particle production
model. Our claim is that FLUKA has already given a more accurate prediction
than the previous calculations. Also in 
the recent work of the Bartol group\cite{bartol_new}, in which some
feature of their hadronic interaction model have been corrected, a neutrino
flux very close to our predictions is now obtained for high cut--off
sites, like Super--Kamiokande.
It is likely that in the case of Super--Kamiokande analysis
the differences between fluxes are 
in practice obscured by the assumed systematic uncertainties.
However, in view of possible future improvements, 
it is important to clarify the differences among the existing
predictions.
The effects due to the 3-D treatment of the shower, first 
presented in ref.\cite{fluka3d}, have been reproduced by
other groups\cite{lipgeo,honda2001} and are by now accepted as the correct
prediction. The uncertainties in the neutrino rates are instead less
clear, but they can be attributed to three independent sources:
i) the primary cosmic ray spectrum, ii) the hadronic
interaction model and iii) the neutrino--Nucleus cross sections.
The first goal of this paper (section \ref{sec:muons})
is to provide further evidence that the
particle yields predicted by the FLUKA model\cite{fluka3d,now2000} are in the
right range. 
As a second step, we consider the
dependence of the predicted $\nu$ flux on the choice of primary spectrum
(section \ref{sec:spect}).
The fundamental topic of neutrino cross-sections
is instead postponed to a future work.
In the conclusions we summarize the arguments, advancing some criticism
about other new attempts of $\nu$ flux calculations.

\section{New benchmarks for the FLUKA predictions}
\label{sec:muons}
FLUKA is constructed upon a set of theoretically inspired models of
hadronic interactions, adjusting parameters only on the basis of 
results coming from accelerators. 
Cosmic ray data can however be used to check their quality. Recently
two remarkable results have been achieved in this field.
The first one is the reproduction of the features of the primary proton flux as
a function of geomagnetic latitude as measured by AMS\cite{zuccon},
thus showing that both the production of secondary nucleons and the
geomagnetic effects and the overall geometrical 
description of the 3--D setup are well under control. 
The same work also shows that
the fluxes of secondary $e^+$ and $e^-$ measured at high altitude are
reproduced, and this is instead more directly linked to
the yield of mesons produced in primary interactions.
The second achievement is the good reproduction of the muon data in
atmosphere as measured 
by the CAPRICE experiment\cite{caprice}, both at ground level and at different
floating altitudes\cite{caprice_fluka}, when starting from the same primary
flux (Bartol fit)\cite{bartol} used to generate atmospheric neutrinos.
The fluxes of atmospheric muons are strictly related to the neutrino ones,
because almost all $\nu$'s are produced either in association, with, or in the
decay of $\mu^\pm$. 
The level of agreement reported in ref.\cite{caprice_fluka}, without any
adjustment of the model, allows
to conclude that, the normalization of the $\nu$ fluxes predicted by
FLUKA is in the right range. This is true for a choice of primary spectrum
constrained by the more recent data, in particular
from AMS\cite{ams} and BESS\cite{besspr}. The quoted primary flux of
ref.\cite{bartol} is already 
in good, although not yet optimal, agreement with these data.
Furthermore, the agreement exhibited by the FLUKA simulation separately for
muons of both charges gives confidence on
the simulated yields of $\pi^+$ and $\pi^-$ production in air. This is one
of the main sources of uncertainty in the prediction of the $\mu/e$
ratio. The decay chain of $\pi^+$ eventually results in $\nu_e$, while the
decay of $\pi^-$ 
generates $\nubar_e$. Due to the different interaction cross section this
affects the experimental event rates.

\section{Neutrino Fluxes and Primary Spectrum}
\label{sec:spect}
\begin{figure}[hbtp]
\begin{center}
\includegraphics[width=12cm]{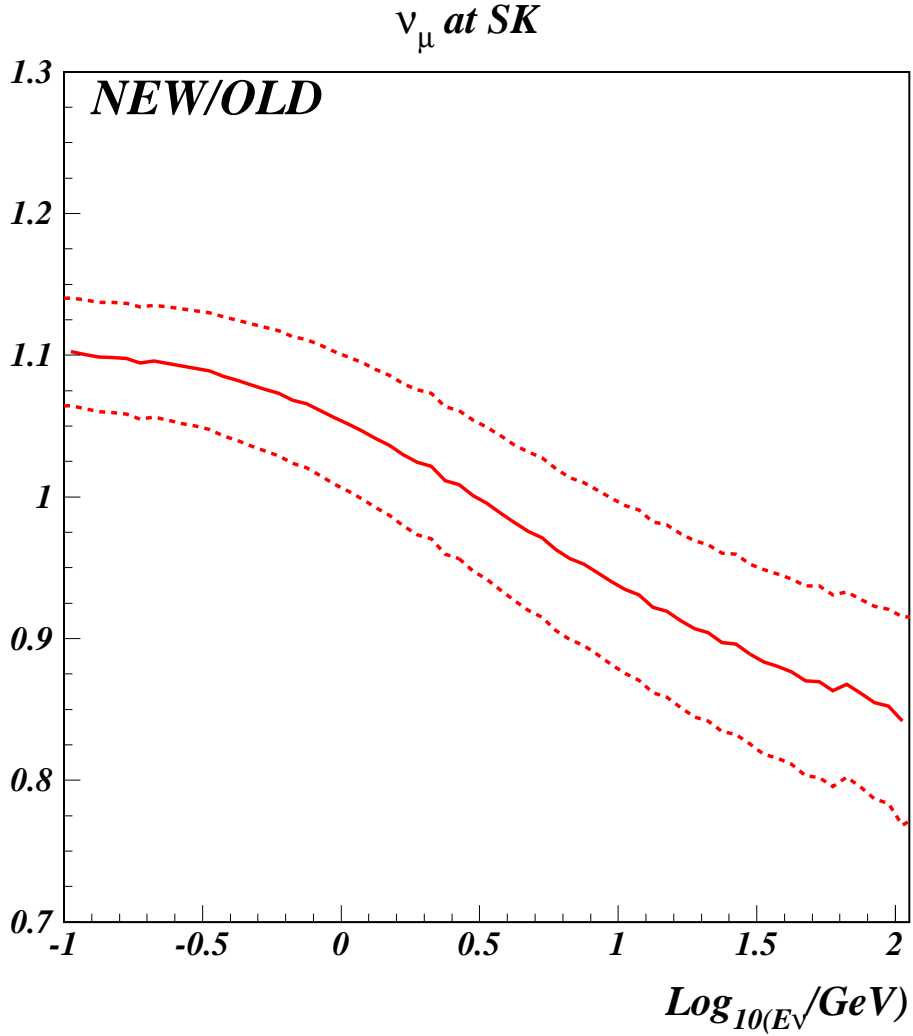} 
\vspace{-1cm}
\caption{Ratio of the computed
$\nu_\mu$ flux at Super--Kamiokande as resulting from the new Bartol fit
with respect to the old fit, as a function of $\nu$ energy. The uncertainty
band is also shown.
\label{fig:gs1}}
\end{center}
\end{figure}

For a given shower model, the level of
agreement in a comparison to measured particle fluxes in atmosphere 
depends on the parameterization of the primary spectrum.
The primary choice originally used in ref.\cite{hkkm}
would force in FLUKA a too high normalization of the flux, since it is
includes the data points taken from the compilation of
Webber et al.\cite{webber}, now excluded by recent experiments.
On the contrary, a constraint based on the primary data from
CAPRICE\cite{caprice_prim} 
would not allow in FLUKA a satisfactory level of agreement with muon
fluxes. As already mentioned,
the Bartol fit used so far is not completely satisfactory. 
The scientific community is trying to establish a common
reference for the primary flux to be used in all computations.
An attempt in this direction has been proposed in ref.\cite{bartol_prim},
where the statistical errors of measurements and the systematic differences
between experiments are used to establish a band of uncertainty
around the average result of this new fit. Contributions heavier than
primary protons are also considered. A similar attempt has been
proposed in ref\cite{honda2001}. 
We have used the new Bartol fit of primaries to recalculate the FLUKA
$\nu$ fluxes. As an example, in  Fig. \ref{fig:gs1} we plot the
ratio of the computed
$\nu_\mu$ flux for the site of Super--Kamiokande as resulting from the
new Bartol fit with respect to the old one, as a function of $\nu$ energy.

The new parameterization brings to an increase of 
neutrino fluxes in the Sub--GeV range, while a depletion occurs in the
Multi--GeV range. This behavior is still maintained when event rates are
considered. Integrating in energy, 
the average prediction for the
atmospheric neutrino event rates increases, approximately for all flavors,
by $\sim$5\% with respect to the original evaluation, with an
uncertainty (only due to the primary spectrum) of $\sim \pm$4\%.

\section{Conclusions}
\label{sec:concl}
The absolute normalization of atmospheric neutrino
fluxes is still a subject of investigation.
The results summarized in this paper support the conclusion that
the normalization of the FLUKA results should be close to reality and that
the difference  with respect to the original Bartol and HKKM results cannot
be attributed to an insufficient particle production yield in the shower
development in the FLUKA model. 
The knowledge of the primary spectrum is still an important issue, but
the convergence towards a common reference, constrained by the AMS and BESS
data allows to reduce the related uncertainty ($\pm$ $\sim$5\%).
After this study we can
also conclude that the agreement of FLUKA predictions to
Super--Kamiokande\cite{Totsuka} data will certainly improve if the updated
parameterization of the proton and He fluxes are used. This should be more
evident for the Sub--GeV events.
This does not mean that the questions about the absolute values of neutrino
cross-sections and the hadronic
interaction descriptions are answered. In particular we remind that 
hadron--Nucleus and Nucleus--Nucleus interactions in  
the energy range 1$\div$30 GeV are still insufficiently known to allow
the construction of a fully reliable model.
In any case it seems difficult that the overall 
uncertainties are larger than 15\%$\div$20\%. Therefore, 
in our opinion, the results on neutrino flux of ref.\cite{naumov}, starting
from a primary spectrum very similar to the one here considered, seem too
low in the Sub--GeV region, not being 
compatible with the corresponding muon fluxes in atmosphere, unless
significant systematic error in the data at high altitude are
assumed. Furthermore, after our tests on direct Nucleus--Nucleus interaction,
by means of an interface to the DPMJET
code\cite{dpmjet}, we cannot share their conclusion about the existence of
a significant bias introduced by the use of the superposition model.
As far as the $\nu$ fluxes proposed in ref.\cite{plyask}, we believe that
they are probably affected by some severe error (see also
ref.\cite{lip2001}). Beyond other considerations, those
results would imply, without any supporting evidence, that
either the experimental data on atmospheric neutrinos are strongly biased
or that the present knowledge of neutrino cross sections are wrong by an
unexplained large factor.

\end{document}